# YanTian: An Application Platform for AI Global Weather Forecasting Models

Wencong Cheng[1], Jiangjiang Xia[2*], Chang Qu[2], Zhigang Wang[1], Xinyi Zeng[2], Fang Huang[1], Tianye Li[2]

[1]Beijing Aviation Meteorological Institute, Beijing 100085, China

[2]Key Laboratory of Regional Climate-Environment for Temperate East Asia & Center for Artificial Intelligence in Atmospheric Science, Institute of Atmospheric Physics, Chinese Academy of Sciences, Beijing 100029, China

[*]Correspondence author: xiajj@tea.ac.cn

Contributing authors：emailtocheng@163.com; quchang@tea.ac.cn; wzg_wang@163.com; zengxy@tea.ac.cn; fangfangzuitian@163.com; lity@tea.ac.cn

**Abstract:** To promote the practical application of AI Global Weather Forecasting Models (AIGWFM), we have developed an adaptable application platform named "YanTian". This platform enhances existing open-source AIGWFM with a suite of capability-enhancing modules and is constructed by a "loosely coupled" plug-in architecture. The goal of "YanTian" is to address the limitations of current open-source AIGWFM in operational application, including improving local forecast accuracy, providing spatial high-resolution forecasts, increasing density of forecast intervals, and generating diverse products with the provision of AIGC capabilities. "YanTian" also provides a simple, visualized user interface, allowing meteorologists easily access both basic and extended capabilities of the platform by simply configuring the platform UI. Users do not need to possess the complex artificial intelligence knowledge and the coding techniques. Additionally, "YanTian" can be deployed on a PC with GPUs. We hope "YanTian" can facilitate the operational widespread adoption of AIGWFMs.

# 1   Introduction

In recent years, studies of AI Global Weather Forecasting Models (AIGWFM) have become an emerging direction in the development of atmospheric science. There have been important technological breakthroughs in the application of AI in weather forecasting. Huawei, Google DeepMind, Fudan University, Shanghai AI Laboratory, NVIDIA and other organizations have launched a series of AIGWFMs, such as Pangu-weather (Bi *et al*., 2023), GraphCast (Lam *et al*., 2022), Fengwu (Chen *et al*., 2023a), Fuxi (Chen *et al*., 2023c), FourCastNet (Pathak *et al*., 2022) , and others. Previous evaluation studies of these AIGWFMs show that the forecasting capabilities of the data-driven AI weather (0-14 days or so) forecasting models are comparable to those of numerical weather prediction models (Bouallègue *et al.*, 2023; Cheng *et al.*, 2023), attracting much attention in the meteorological community.

Furthermore, it is widely acknowledged that the AIGWFMs have notable advantages over traditional numerical weather prediction models. For example, while numerical weather prediction models require substantial computational resources, AIGWFMs use significantly less computational resources for inference forecasting, allowing them to be deployed even on small personal computers, thus expanding the scenarios of weather forecasting applications. By using AIGWFMs, ocean going vessels or weather forecasting departments in remote areas, which are generally unable to access (or purchase) abundant weather data due to data transmission limitations and thus do not have the self-production capability of weather forecasts, can also make quantitative weather forecasts based on limited global initial data and local observation data to meet their forecasting needs. In addition, AIGWFMs are fast, and most of them can generate short- to medium-term weather forecasts for the next 10 more days within 1 to 2 minutes, and can therefore be used to generate multi-member ensemble forecasts and form probabilistic forecasts for extreme weather as a supplement to the results of deterministic weather prediction models.

Leveraging these advantages, the AIGWFMs have shown significant potential for the development of meteorology. However, several issues need to be addressed when applying AIGWFMs in operational applications: most AIGWFMs use ERA5 reanalysis products for training, leading to notable homogeneity in forecasting capabilities, including resolution, output product types, and forecast intervals. Given the current AIGWFMs forecasting capabilities have been comparable to the traditional numerical weather prediction, further marginal improvements in accuracy for specific forecast elements by designing and training new models would require substantial additional costs and the benefit is limited. Issues that remain unsolved including improving local forecast accuracy, downscaling spatial resolution, increasing forecast interval density, and generating diverse products for operational applications (such as aviation meteorology and renewable energy applications). To address these issues, some institutions are conducting their research on AIGWFMs, including developing AI assimilation (Chen *et al.*, 2023b), sub-seasonal forecast (Chen *et al.*, 2023d), downscaling (Mardani *et al*., 2023), extreme event forecast (Zhong *et al*., 2023), and so on, aiming at achieving more accurate, longer-range, higher resolution, and more product types of weather forecasts. However, these models with 'specific function' are either constructed for specific regions, specific elements, specific models, or are not yet open sourced for public use and cannot be widely applied at present. Overcoming the limitations imposed by the general used training reanalysis dataset, reconstructing a global, higher spatial and temporal resolution training dataset with more elements, and developing new, more capable AIGWFMs on top of it will require orders of magnitude more investment (algorithms, computational source, data) and time. Achieving the work is a considerable challenge.

Against this background, the present study constructs the "YanTian" (means "forecast the weather" in Chinese) application platform on the basis of AIGWFMs that have been trained and released (open source) so far. "YanTian" features a top-level design that combines the basic AIGWFMs with a series of capability-enhancing embedded modules, constructing a practical forecasting application platform by a "loosely coupled" architecture. The primary goal of "YanTian" is to address the limitations of the existing open-source basic AIGWFMs in operational applications without spending too much resources on re-training the basic AIGWFMs, so as to improve local forecast results accuracy, provide spatial high-resolution forecasts, increase forecast time interval density, and generate diverse products with the provision of AIGC capabilities.

In addition, the system has the following two characteristics:

1. Users do not need to possess the complex artificial intelligence knowledge and coding techniques, and they can apply the basic and extended forecasting capabilities of the platform by simply configuring the visual interface (primarily through selecting several forecast-related parameter options on the platform's UI);

2. It can be deployed and run on a single PC, requiring only hardware configurations that can support the operation of the basic AIGWFMs. For example, if the Pangu-weather forecast model is used as the basic model, a GPU with more than 14 GB of memory (e.g. NVIDIA RTX 4090, RTX 3090, RTX 3080) is sufficient.

We believe these two characteristics represent the general operational pattern of quantitative weather forecasting capabilities in the age of AIGWFMs. Specifically, many small and medium-sized meteorological service providers will develop customized and refined weather forecasting capabilities through combination of capability-enhancing modules and the existing AIGWFMs. We hope that the "YanTian" Application Platform can facilitate the operational widespread adoption of AIGWFMs, and achieve practical application in multiple weather-related scenarios.

## 2 "YanTian" Application Platform

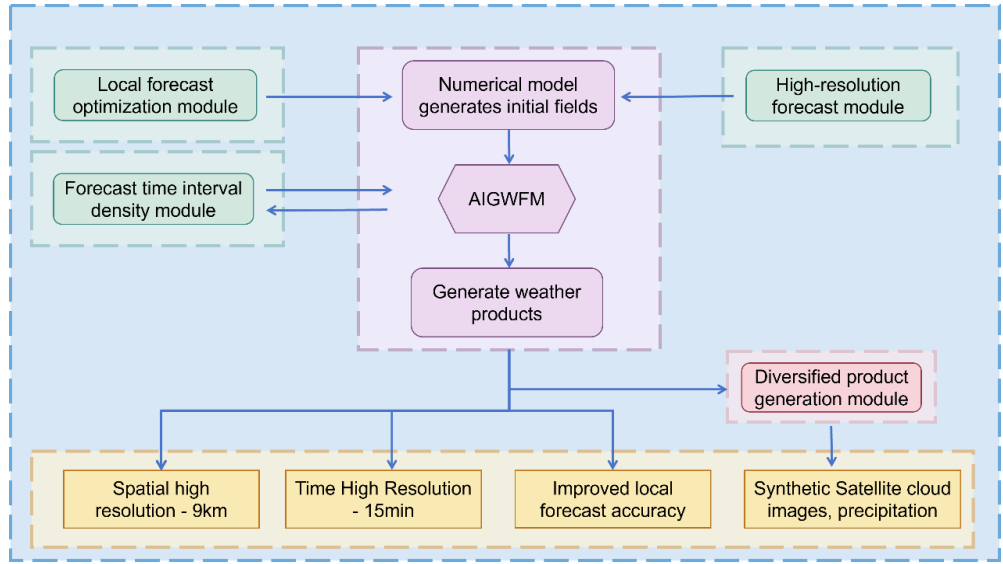

Figure 1. Architecture of the "YanTian" application platform

The architecture of "YanTian" is shown in Figure 1. It is designed based on the concept of openness and scalability, and currently includes four application modules: local forecast optimization module, high-resolution forecast module, forecast time interval density module, diversified product generation module. Each module has a well-designed core algorithm to support it. And an application platform software that supports visual interactive operation.

**2.1 Local Forecast Optimization Module**

2.1.1 Module overview

**Module Function:** The Local Forecast Optimization Module trains a deep neural network by using a Mask operation with sparse station data to correct the gridded forecast data. This process is expected to learn from sparse data, but to produce a dense forecast and then uses the corrected forecast grid data as the initial input data for subsequent forecasts by the AIGWFMs. This approach improves the accuracy of the local forecasts of the subsequent times, effectively allowing the AIGWFM to integrate local station observation data.

**Rationale:** The initial data for AIGWFM are typically based on numerical weather prediction analysis data. Once the forecasting process initiates, these models cannot utilize subsequently updated regional observational data. Our previous research (Cheng *et al.*, 2023) showed that fusion regional data which are more accurate into the initial input data of the AIGWFM, referred to as "local assimilation", can improve the accuracy of local forecasts.

**Input and Output:** In the inference stage, the local forecast optimization module densifies station observations into regional grid-calibrated results, which are then fused with existing forecast products to form a new initial data. This new initial data is then fed back into the AIGWFM, resulting in more accurate forecast products for subsequent time steps.

2.1.2 Module Principles

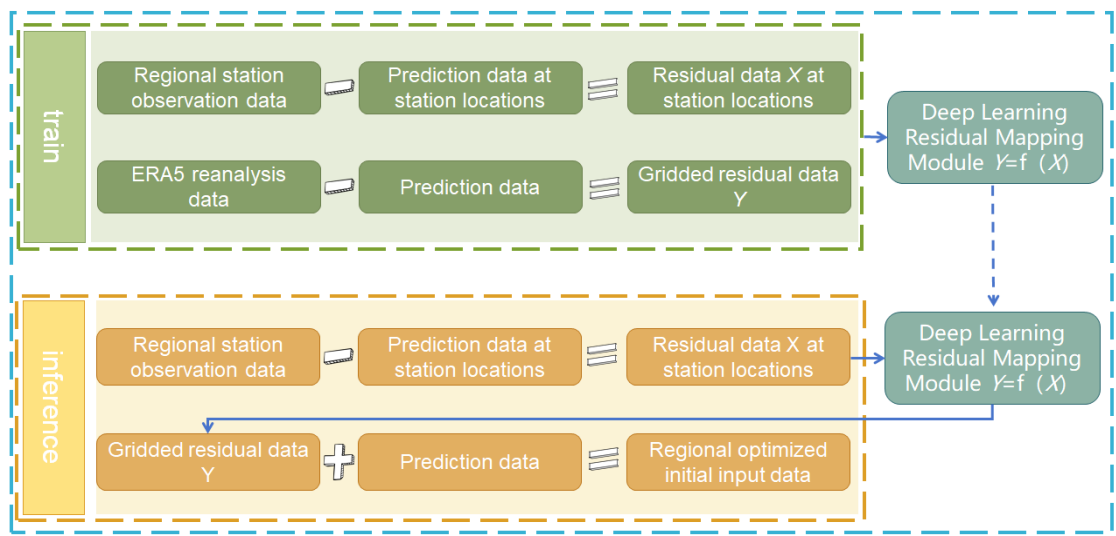

Figure 2. Architecture of the local forecast optimization module

As shown in Figure. 2, in the training stage, regional station observation data is used to subtract the prediction data at station locations, to get the residual data X at station locations. Then, ERA5 reanalysis data is used to subtract the prediction data to get the gridded residual data Y. A deep neural network is then trained to establish the mapping relationship between X and Y, i.e., Y=f(X). This process enables a densification from station observation values to grid values. The definition of densification is referenced from MetNet3 (Andrychowicz *et al*., 2023). For example, the station observation data are obtained from the China Meteorological Administration.

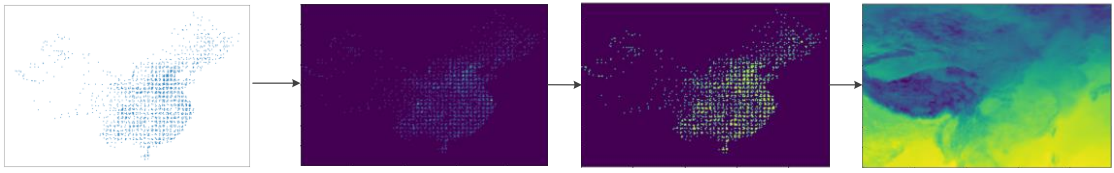

Figure 3. Densification from station observation values to grid values. a) positions of the observation stations from China Meteorological Administration; b) projection of stations onto a 0.25 ° grid point, with darker colors indicating higher stations density within each grid point; c) meteorological observation data at the grid points, i.e. initial field data at the station locations, taking 2m temperature as an example; d) grid values after densification, i.e. regional optimized initial data.

In the inference phase, the regional station observation data are used to subtract the prediction data at station locations to obtain the residual data X at station locations. X is used as an input of the deep learning residual mapping module f(X) to obtain the gridded residual data Y. Y is then summed with the prediction data to obtain the regional optimized initial input data. Finally, the regional optimized initial input data are fused into the global prediction data as the input data of AIGWFM to generate the following time steps forecast products

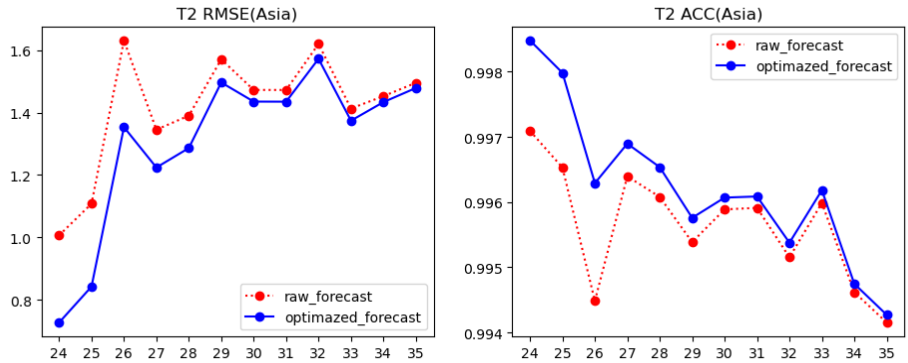

Figure 4. Effect of optimizing the regional initial data as an input of the AIGWFM using stations data in the regional area as shown in Figure. 3.

For example, we fuse the observational information from the 24-h forecast output of the Pangu-weather model using the aforementioned method, and then use it as the input data of the Pangu-weather model. The results in Figure 4 show that the local forecast optimization using the information of the regional stations can significantly improve the accuracy of the subsequent 12-h forecasts of the region.

This finding confirms the conclusions of our previous related study (Cheng *et al*., 2023).

**2.2 High-resolution forecast module**

2.2.1 Module overview

**Module Function:** The high-resolution forecast module is based on the idea of 'Pixel Shifting' technology (Allen *et al.*, 2005) of the video projectors, combined with the diffusion algorithm to support the initial data of 9km resolution (about 0.08° latitude/longitude resolution) as the input of the AIGWFMs, and generate corresponding high-resolution forecast products.

**Rationale:** Current AIGWFMs are mainly trained using ERA5 reanalysis data (0.25° geographical resolution). Both the inputs and outputs of these models are limited by this resolution, which usually does not meet the common practical operational demands. Although some global operational numerical weather prediction models have already increased the resolution to 9km or better currently (e.g., ECMWF IFS High-Resolution Operational Forecasts https://rda.ucar.edu/datasets/ds113.1/dataaccess/), existing AIGWFMs cannot utilize such high-resolution data as the initial inputs to produce high-resolution forecasting outputs. Directly downscaling would result in significant loss of the initial information. Re-training a new AIGWFM to support high-resolution inputs and generate high-resolution forecast products requires substantial resources. Therefore, we designed a high-resolution forecast module to leverage the capabilities of existing AIGWFM to support the input and output with high-resolution.

**Input and output:** the input for the module is the analysis data with a 0.08° geographical resolution. As the common elements of the currently AIGWFMs, the input data include four surface meteorological elements: 2-meter temperature (2T), 10-meter u-component of wind (10U), 10-meter v-component of wind (10V), and mean sea level pressure (MSL); Thirteen pressure levels (50, 100, 150, 200, 250, 300, 400, 500, 600, 700, 850, 925, and 1000 hPa) with five variables per level: u-component of wind (east-west component), v-component of wind (north-south component), temperature, specific humidity, and geopotential height. The module's output matches its input.

2.2.2 Module Principles

In this module, we propose a 'mean-variance' based method to forecast the high resolution products. We take the "Pixel Shifting" method as the 'mean' component and the conditional diffusion method as the 'variance' component to generate the high resolution forecast products.

'Mean' component: In the domain of the video projector and camera technology, the "Pixel Shifting" technique enables the generation of high-resolution images using a set of lens that originally only supports lower resolutions. Inspired by this downscaling technique, the block operations are performed on the high-resolution initial data according to the resolution of the basic AIGWFM. For each specific location within the blocks, a forecast product is generated using the basic AIGWFM, and then these low-resolution outputs are combined to generate the coarse high-resolution forecasts('mean' component) as the first step products. The similar method was also adopted by Fengwu-GHR (Han *et al.*, 2024). The forecasting process is illustrated in Figure 5.

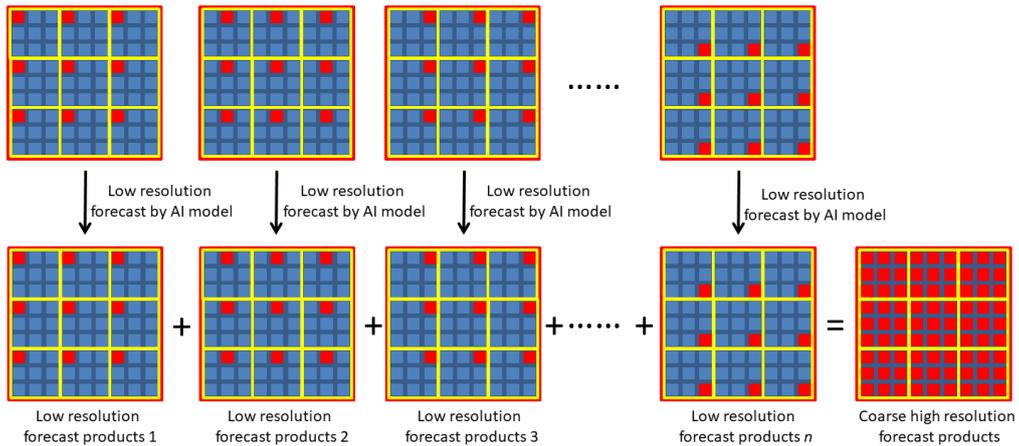

Figure 5. "Pixel Shifting" in the High Resolution Forecast Module of the "Yantian"

'Variance' component: Through the above process, we keep the whole information of the initial high-resolution data to the forecast products, and solely utilizing the capabilities of the basic AIGWFM. However, since the basic AIGWFM has not been optimized for local meteorological information, its outputs do not reflect local meteorological characteristics. Based on the coarse high-resolution forecast product generated by 'Pixel Shifting', we draw inspiration from the Corrdiff (Morteza Mardani *et al*., 2023), which is the core algorithm used in NVIDIA's Earth-2 system. The Corrdiff algorithm separates the target product into a 'mean' component and a 'variance' component, which are calculated separately before being combined. The "mean" component uses a U-Net architecture to capture trend information, while the 'variance' component utilizes a diffusion model to acquire local information. Similarly, in our high-resolution forecast scenario, we consider the global prediction results obtained through the 'Pixel Shifting' approach as the 'mean' part of the high-resolution prediction product. The difference between the 'mean' and the ground truth serves as the 'variance' target. A conditional diffusion model is introduced to generate the local 'variance' target, which is then added to the 'mean' component to produce the final forecast products, fusing the global and local meteorological features. This process is illustrated in Figure 6. An example of the high-resolution forecasting product is shown in Figure 7.

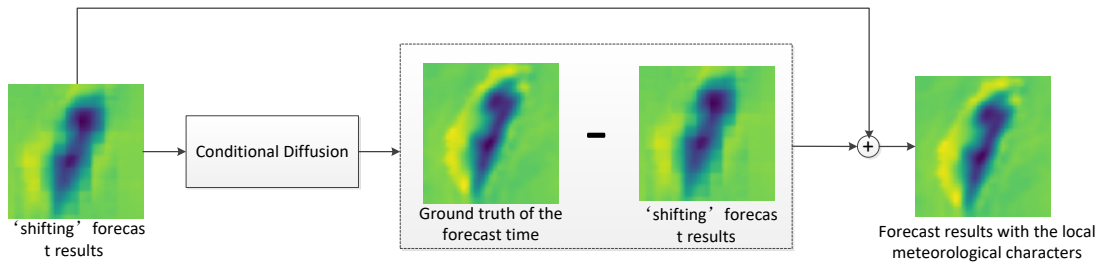

Figure 6. Process for obtaining the high-resolution prediction product based on the diffusion model.

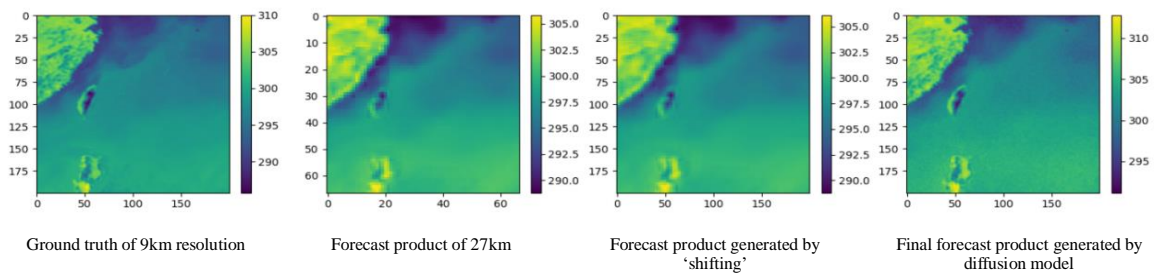

Figure 7. An example of the high-resolution forecast product (2m temperature)

**2.3 Forecast time interval density module**

2.3.1 Module overview

**Module function:** The forecast time interval density module primarily utilizes bi-directional optical flow interpolation and the aforementioned local observational data assimilation techniques to enhance the time interval of existing AIGWFMs from hourly to 15-minute.

**Rationale:** Current AIGWFMs are primarily trained using ERA5 reanalysis data (with an hourly time resolution), hence the minimum forecast interval of AIGWFMs' output is also one hour. This does not meet the normal requirements of many operational scenarios, especially those needing high spatial resolution forecasting products, where the forecast interval needs to be further densified to satisfy operational usage requirements. To address this, we have designed the forecast time interval density module to be compatible with existing AIGWFMs, which can integrate updated observational data to produce forecast products at 15-minute intervals.

**Input and output:** The input of the forecast time interval density module is the common initial data of the basic AIGWFM, and the output is the forecast products with 15-minute time intervals.

2.3.2 Module Principles

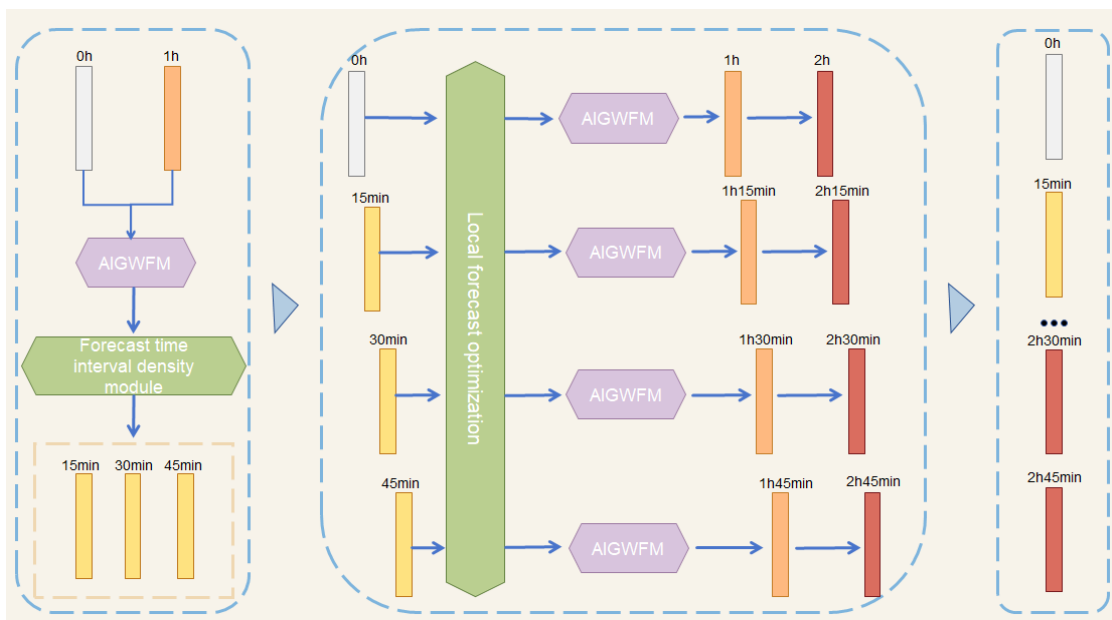

Figure 8. The Forecast Interval Density Module of the "Yantian"

As shown in Figure 8, the implementation of this module consists of three steps:

Temporal Interpolation: The initial data at time 0 and the forecast products at the 1-hour are used as the inputs. This module outputs data at 15-minute, 30-minute, and 45-minute intervals. The forecast time interval density module employs bi-directional optical flow interpolation technique, which is commonly used in video interpolation.

Observational Data Fusion: Observation data obtained every 15 minutes is assimilated into the interpolated data using the aforementioned local forecast optimization method. This step enhances the accuracy of the interpolated data.

Forecast Update: Updated data at 15-minute, 30-minute, and 45-minute intervals are then used as initial inputs for the AIGWFM. Forecast products are generated for future times such as 1 hour and 15 minutes, 1 hour and 30 minutes, 1 hour and 45 minutes, 2 hours, and 2 hours and 15 minutes, etc. Figure 9 shows an example of the forecast time interval density.

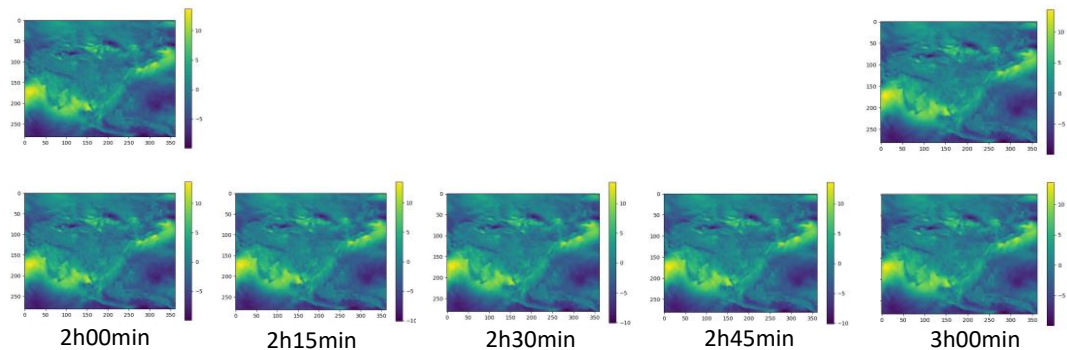

Figure 9. An example of Forecast Time Interval Density (results of U-wind field over east Asia from the 2 to 3 lead hour, first row shows original 1-hour intervals, second row shows 15-minute intervals)

## 2.4 Diversity Product Generation Module

2.4.1 Module overview

**Module function:** The diversified product generation module utilizes the outputs of the AIGWFM to generate a variety of weather forecast products, such as precipitation forecast products and synthetic satellite cloud images.

**Rationale:** While the current AIGWFMs' forecast elements cover the common factors used in forecast products evaluating, similar to those in numerical weather prediction models, they can serve as a measure of forecasting capability. However, the range of the products is still relatively limited, lacking some derivative forecast products required by some weather-related industries. For example, the outputs of AIGWFMs typically do not include precipitation forecast products, nor do they provide specific elements of interest such as cloud positions for the aviation meteorology industries. Thanks to the advance development of AIGC (Artificial Intelligence Generated Content), we can derive various weather-related forecast products based on the output of the AIGWFMs. For the first step, we have developed forecasting precipitation products and synthetic cloud image generation capabilities in "Yantian".

**Input and Output:** In the inference stage, the input of the module is the output of the AIGWFMs and a few additional elements, and the output of the module is precipitation products, or synthetic cloud images, and so on.

2.4.2 Module Principles

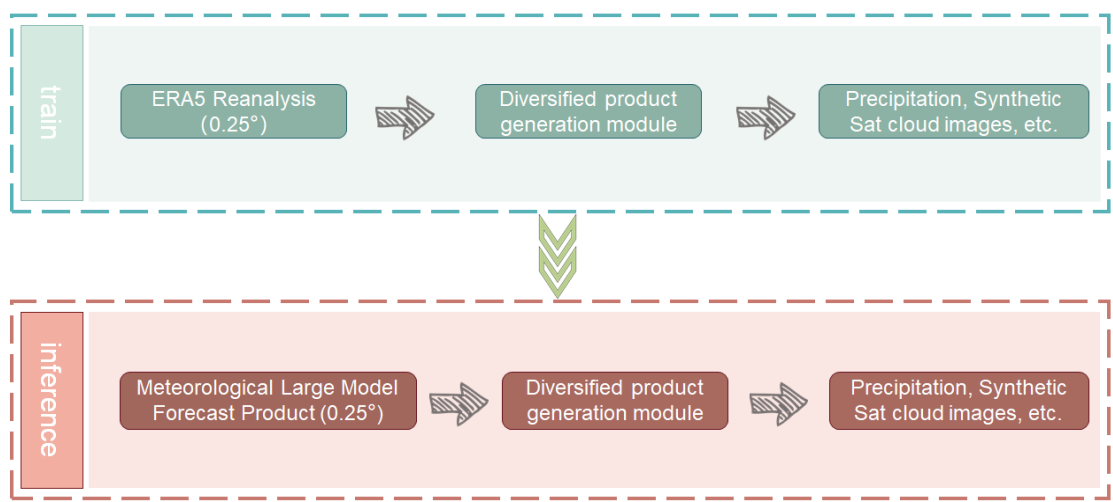

Figure 10. Diversified product generation module of "Yantian"

Training Phase: During the training phase, ERA5 data corresponding to the types and levels of the outputs by AIGWFMs and a few additional elements are used as inputs for the diversified product generation module. Then we train multiple deep learning models, such as the precipitation model and the synthetic cloud image model as the sub-modules. The outputs include precipitation forecasts, synthetic cloud images, and other derived forecast products.

Inference Phase: During the inference phase, the deep learning diversified product generation module uses the outputs of the AIGWFM and a few additional elements as inputs and then invokes the specific sub-modules to obtain a diversified set of meteorological forecast products.

A brief description of each module is given below:

**(1) Precipitation Sub-module**

The meteorological elements input of the precipitation sub-module includes four atmospheric variables: 2-meter temperature (2T), 10-meter u-component of wind (10U), 10-meter v-component of wind (10V), and mean sea level pressure (MSL); Six geographical information variables: longitude, latitude, altitude, angle of sub-grid scale orography, slope of sub-grid scale orography, and land-sea mask, which represent geographical information and the impact of land-sea differences on precipitation. Thirteen pressure levels (50, 100, 150, 200, 250, 300, 400, 500, 600, 700, 850, 925, and 1000 hPa) with five variables per level: u-component (east-west component) of wind, v-component (north-south component) of wind, temperature, specific humidity, and geopotential height.

The output is 3-hour accumulated precipitation amount with a spatial resolution of 0.1°. Precipitation values from ERA5-Land (total precipitation, 0.1°) serve as ground truth labels for land precipitation, while ERA-Single's precipitation values (0.25°) serve as ground truth labels for oceanic precipitation. These are combined to form the precipitation ground truth for the target region (0°-60°N, 70°-140°E).

This sub-module is based on the Swin-Transformer structure, utilizing self-attention mechanisms and patch processing to establish global connections between multi-scale meteorological elements, enabling more accurate precipitation forecasting results. An example of the performance of the precipitation sub-module is shown in Figure 11.

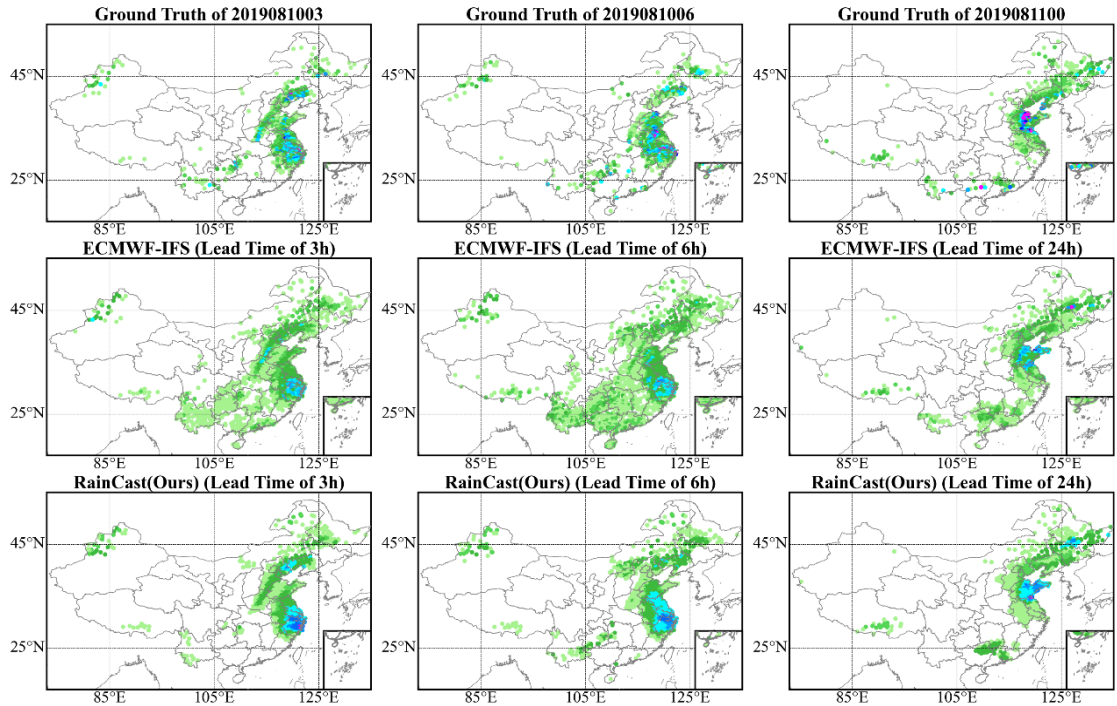

Figure 11. An example illustrates the results of the precipitation sub-module. The first row shows the observed values, the second row displays the ECMWF-IFS forecast results, and the third row presents the forecast effects of the precipitation sub-module.

**(2) Synthetic Satellite Cloud Images Sub-module**

The meteorological elements input into the synthetic cloud image sub-module includes four atmospheric variables: 2-meter temperature (2T), 10-meter u-component of wind (10U), 10-meter v-component of wind (10V), and mean sea level pressure (MSL); Thirteen pressure levels (50, 100, 150, 200, 250, 300, 400, 500, 600, 700, 850, 925, and 1000 hPa) with five variables per level: u-component of wind (east-west component), v-component of wind (north-south component), temperature, specific humidity, and geopotential height.

The output is the synthetic images corresponding to the FY-4B meteorological satellite's infrared 12 channels (band: 10.3-11.3 μm) and visible light channels 1, 2, and 3, with a spatial resolution of 4 km. The spatial coverage is from 10° to 50° N latitude and 90° to 130° E longitude.

This module is based on a generative adversarial network (GAN) integrated with an attention mechanism. For detailed methods, see the references Cheng *et al*. (2020, in Chinese; and 2022).

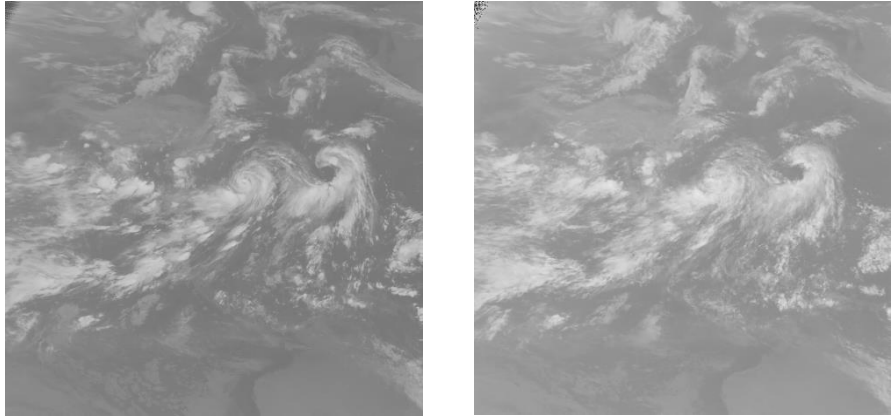

a) Real satellite cloud image     b) Synthetic satellite cloud image

Figure 12. Illustrative Effect of the Synthetic Satellite Cloud Images Sub-module (FY-4B Meteorological Satellite Cloud Image)

## 3   Platform Software

To enable operational meteorologists to conveniently access AIGWFMs locally and enhance localized forecasting capabilities, we have developed "YanTian" AIGWFM Application Platform software. This platform software integrates various invoke interface of the basic AIGWFMs with local forecast optimization module, high-resolution forecast module, forecast time interval density module, and diversity product generation module. Users can invoke the functions of the AIGWFMs and the additional modules through a visual parameter configuration user interface.

For ease of AI algorithm integration, the back-end part of the platform is implemented in Python, encapsulating the algorithm functions as the callable services using the 'FastAPI' open-source framework. The front-end is designed as a plugin interactive interface using the 'VUE3' framework for services invocations. This platform supports cross-platform deployment and is compatible with both Windows and Linux environments.

**3.1 Parameter Settings and Usage of the Basic Forecast Function**

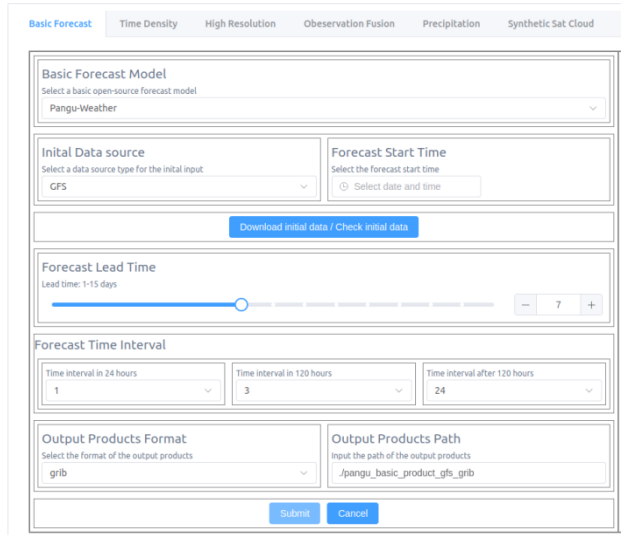

Figure 13. User interface of the basic forecast function in "YanTian" platform software

Through the "YanTian" Platform software, users can configure parameters and perform forecasting inference by the basic AIGWFM via a visualized UI interface. The basic forecast models include various open-source AIGWFMs (users need to manually download the required AIGWFMs, as the application platform software does not pre-install any basic AIGWFM).

The basic forecast function interface includes the following configurable parameters:

（1）Basic Forecast Model Selection

Various existing open-source AIGWFMs can be selected for configuration, such as Pangu-weather, GraphCast, Fuxi, etc.

（2）Initial Data Source Selection

Initial meteorological forecast data like the American GFS, ECMWF IFS, or the CMA-GFS can be chosen as input for the AIGWFM. Built-in decode modules are available to directly read data files as model inputs.

（3）Forecast Start Time Selection

Users can select the start time for the forecast, choosing pre-downloaded initial data on the local machine as model inputs. It also supports direct downloading of initial data for certain data source (such as American GFS forecasts) at the selected time step.

（4）Forecast Lead Time Selection

Users can choose forecast lead time to generate forecast products for the range of 1 to 15 days.

（5）Forecast Time Interval Selection

Different forecast intervals can be selected within the forecast lead time, such as hourly for the first 24 hours, every three hours for the next 120 hours, and every 24 hours after 120 hours.

（6）Data Output Format Selection

The application platform wraps the outputs of AIGWFMs and provides forecast product data in GRIB, NetCDF (NC), or NumPy (NPY) formats.

（7）Output Path Selection for Products

Users can specify their own output data paths for forecast products.

After completing the configuration of the forecast model parameters, users can click the "Submit" button to execute the AIGWFM forecasting according to the set parameters. The results will be generated and saved to the specified directory. The samples of intermediate products and the progress percentage information during the generation will be shown in the right part of the UI.

**3.2 Parameter Settings and Usage of the Local Forecast Optimization Function**

Figure 14. User interface of the Local forecast optimization function in "YanTian" platform software.

By running the local forecast optimization module, the local meteorological initial data is optimized to improve regional forecasts. The local forecast optimization

function mainly includes the following optional parameters:

（1）Basic Forecast Model Selection

Various existing open-source AIGWFMs can be selected for configuration, such as Pangu-weather, GraphCast, Fuxi, etc.

（2）Initial Data Selection

Initial meteorological forecast data like the American GFS, ECMWF IFS, or the CMA-GFS can be chosen as input for the function. Built-in decode modules are available to directly read data files.

（3）Forecast Start Time Selection

Users can select the start time for the forecast, choosing stored pre-downloaded initial data fields on the local machine as model inputs. It also supports direct downloading of initial data fields for certain data source categories (such as American GFS forecasts) at the selected time step.

（4）Local Observation Directory Selection

The directory where local observation data is pre-downloaded.

（5）Initial Data completeness Analysis (Check Initial Data)

Analysis of the initial data completeness to determine if there are any missing data based on the above selections.

（6）Forecast Lead Time Selection

Users can choose the forecast lead time for the future range of 1 to 15 days.

（7）Data Output Format Selection

The application platform wraps the outputs and provides forecast product data in GRIB, NetCDF (NC), or NumPy (NPY) formats.

（8）Output Path Selection for Products

Users can specify their own output paths for forecast products.

After completing the configuration of the model parameters, users can click the "Submit" button to execute the local forecast optimization module according to the set parameters. The results will be generated and saved to the specified directory. The samples of intermediate products and the progress percentage information during the generation will be shown in the right part of the UI.

## 3.3 Parameter Settings and Usage of the High-Resolution Forecast Function

Figure 15. User interface of the high-resolution forecast function in "YanTian"

platform software

The platform can generate the forecast products with a 9km resolution, mainly including the following optional parameters:

(1) Basic Forecast Model Selection

Various existing open-source AIGWFMs can be selected for configuration, such as Pangu-weather, GraphCast, Fuxi, etc.

(2) Selection of Single Level Data Directory for initial data with a 9km resolution

Users can select the single level data directory for initial data with a 9km resolution.

（3）Selection of Pressure Level Data Directory for initial data with a 9km resolution

Users can select the pressure level data directory for initial field with a 9km resolution.

(4) Initial Data Completeness Analysis (Check Initial Data)

Analysis of the initial data completeness to determine if there are any missing data based on the above selections.

(5) Forecast Lead Time Selection

Users can choose forecast lead time for the future range of 1 to 15 days.

(6) Forecast Time Interval Selection

Different forecast intervals can be selected within the forecast lead time, such as hourly for the first 24 hours, every three hours for the next 120 hours, and every 24 hours after 120 hours.

(7) Data Output Format Selection

The application platform wraps the outputs and provides forecast product data in GRIB, NetCDF (NC), or NumPy (NPY) formats.

(8) Output Path Selection for Products

Users can specify their own output paths for forecast products.

After configuring the forecast model parameters, users can click the "Submit" button to run the high-resolution forecast module according to the specified parameters. The results will be generated and saved to the specified directory. The samples of intermediate products and the progress percentage information during the generation will be shown in the right part of the UI.

## 3.4 Parameter Settings and Usage of the Forecast time Interval Density Function

Figure 16. User interface of the forecast time interval density function in "YanTian" platform software.

It is designed to generate the 15 minute interval forecast products, as the original time interval is 1 hour. The forecast time interval density module function mainly includes the following optional parameters:

(1) Basic Forecast Model Selection

Various existing open-source AIGWFMs can be selected for configuration, such as Pangu-weather, GraphCast, Fuxi, etc.

(2) Initial Product Directory Selection

Initially, the basic AIGWFM performs forecasting with time interval of one hour, and users can select the output directory for these forecast products.

(3) Forecast Start Time Selection

Users can select the start time for the forecast, choosing stored initial data on the local machine as model inputs. It also supports direct downloading of initial data for certain data source (such as American GFS forecasts) at the selected time step.

(4) Initial Data Completeness Analysis (Check Initial Data)

Analysis of the initial data completeness to determine if there are any missing data based on the above selections.

(5) Selection of the basic forecast time interval

Users can set the forecast time interval for the basic forecast model, typically 1 hour within the first 24 hours.

(6) Forecast time interval selection

Users can select the desired time interval for the encrypted forecast, with a default of 15 minutes.

(7) Forecast Start Time

Users can choose the start time for initiating the forecast.

(8) Density Forecast Lead Time

Users can select the lead time for the forecast, such as 1-3 days.

By clicking the "Submit" button, users can run the forecast time interval density module, which generates forecast products with intervals of 15 minutes or less.

## 3.5 Parameter Settings and Usage of the Diversified Product Generation Function

Figure 17. User interface of the diversified product generation function in the "YanTian" platform software: precipitation (left), synthetic satellite cloud image (right)

The functions of the diversified product generation module mainly include the following optional parameters:

(1) Basic forecast model selection

Various existing open-source meteorological large models can be selected for configuration, such as Pangu-weather, GraphCast, Fuxi, and so on.

(2) Forecast Start Time Selection

Users can select the start time for the forecast, choosing stored initial fields on the local machine as model inputs. It also supports direct downloading of initial fields for certain data categories (such as American GFS forecasts) at the selected time step.

Forecast Lead Time Selection

Users can choose forecast lead time for the future range of 1 to 15 days.

(4) Initial Data Completeness Analysis

Analysis of the initial data completeness to determine if there are any missing data based on the above selections.

(5) Data Output Format Selection

The application platform wraps the outputs and provides forecast product data in GRIB, NetCDF (NC), or NumPy (NPY) formats.

(6) Output Path Selection for Products

Users can specify their own output paths for forecast products.

Ultimately, user can click the "Submit" button to run the diversified product generation module.

## 4  Conclusion

The "Yantian" AIGWFM Application System is currently in a continuous iteration and upgrade process. Guided by the principles of "openness, collaboration, and multiplication," we aim to launch new enhanced modules through multi-party cooperation in the future. We hope to collaborate with more organizations and, inspired by the ecosystems of language and image large models, and standardize the interfaces and usage methods for AIGWFMs. We hope this work can accelerate the practical application of AI weather forecasting, enabling more small and medium-sized weather-related organizations and a wide range of meteorological professionals to benefit from the advancements in AI weather forecasting.

# References

Andrychowicz Marcin, Lasse Espeholt, Di Li, Samier Merchant, Alexander Merose, Fred Zyda, Shreya Agrawal, and Nal Kalchbrenner, 2023: Deep Learning for Day Forecasts from Sparse Observations, arXiv, https://doi.org/10.48550/arXiv.2306.06079

Allen Will, Robert Ulichney 2012: 47.4: Invited Paper: Wobulation: Doubling the Addressed Resolution of Projection Displays. Sid Symposium Digest of Technical Papers, 36(1):1514-1517. DOI:10.1889/1.2036298.

Bi Kaifeng, Lingxi Xie, Hengheng Zhang, Xin Chen, Xiaotao Gu, and Qi Tian, 2023: Pangu-Weather: A 3D High-Resolution System for Fast and Accurate Global Weather Forecast, Nature, https://doi.org/10.1038/s41586-023-06185-3

Bouallègue Zied Ben, Mariana C A Clare, Linus Magnusson, Estibaliz Gascón, Michael Maier-Gerber, Martin Janoušek, Mark Rodwell, Florian Pinault, Jesper S Dramsch, Simon T K Lang, Baudouin Raoult, Florence Rabier, Matthieu Chevallier, Irina Sandu, Peter Dueben, Matthew Chantry, and Florian Pappenberger，2023: The rise of data-driven weather forecasting: A first statistical assessment of machine learning-based weather forecasts in an operational-like context, arXiv, https://doi.org/10.1175/BAMS-D-23-0162.1

Chen Kang, Tao Han, Junchao Gong, Lei Bai, FenghuaLing, JingJia Luo, Xi Chen, Leiming Ma, Tianning Zhang, Rui Su, Yuanzheng Ci, Bin Li, Xiaokang Yang, Wanli Ouyang, 2023a: FengWu: Pushing skillful global medium-range weather forecast beyond 10 days lead, arXiv, https://doi.org/10.48550/arXiv.2304.02948

Chen Kun, Lei Bai, Fenghua Ling, Peng Ye, Tao Chen, Kang Chen, Tao Han, Wanli Ouyang, 2023b: Towards an end-to-end artificial intelligence driven global weather forecasting system, arXiv, https://doi.org/10.48550/arXiv.2312.12462

Chen Lei, Xiaohui Zhong, Feng Zhang, Yuan Cheng, Yinghui Xu, Yuan Qi and Hao Li, 2023c:FuXi: A cascade machine learning forecasting system for 15-day global weather forecast, arXiv, https://doi.org/10.48550/arXiv.2306.12873

Chen Lei, Xiaohui Zhong, Jie Wu, Deliang Chen, Shangping Xie, Qingchen Chao, Chensen Lin, Zixin Hu, Bo Lu, Hao Li, Yuan Qi, 2023d: FuXi-S2S: An accurate machine learning model for global subseasonal forecasts, arXiv, https://doi.org/10.48550/arXiv.2312.09926

Cheng Wencong, Xiaokang Shi, Zhigang Wang, 2020: Creating Synthetic Satellite Cloud Data Based on GAN Method. Journal of System Simulation, 33 (6): 1297-1306. In Chinese.

Cheng Wencong, Yan Yan, Jiangjiang Xia, Qi Liu, Chang Qu, Zhigang Wang, 2023: The Compatibility between the Pangu Weather Forecasting Model and Meteorological Operational Data, arXiv, https://doi.org/10.48550/arXiv.2308.04460

Cheng, Wencong, Qihua Li, Zhigang Wang, Wenjun Zhang, Fang Huang, 2022: Creating synthetic night-time visible-light meteorological satellite images using the GAN method. Remote Sensing Letters, 13(7), 738–745. https://doi.org/10.1080/2150704X.2022.2079016.

Han Tao, Song Guo, Fenghua Ling, Kang Chen, Junchao Gong, Jingjia Luo, Junxia Gu, Kan Dai, Wanli Quyang, Lei Bai, 2024: FengWu-GHR: Learning the Kilometer-scale Medium-range Global Weather Forecasting, arXiv, https://doi.org/10.48550/arXiv.2402.00059

Lam Remi, Alvaro Sanchez-Gonzalez, Matthew Willson, Peter Wirnsberger, Meire Fortunato, Alexander Pritzel, Suman Ravuri, TimoEwalds, FerranAlet, Zach Eaton-Rosen, Weihua Hu, Alexander Merose, Stephan Hoyer, George Holland, Jacklynn Stott, Oriol Vinyals, ShakirMohamedand Peter Battaglia, 2022: GraphCast: Learning skillful medium-range global weather forecasting, arXiv, https://doi.org/10.48550/arXiv.2212.12794

Mardani Morteza, Noah Brenowitz, Yair Cohen, Jaideep Pathak, Chieh-Yu Chen, Cheng-Chin Liu, Arash Vahdat, Karthik Kashinath, Jan Kautz, Mike Pritchard, 2023:Residual Diffusion Modeling for Km-scale Atmospheric Downscaling, arXiv, https://doi.org/10.48550/arXiv.2309.15214

Pathak Jaideep, Shashank Subramanian, Peter Harrington, Sanjeev Raja, Ashesh Chattopadhyay, Morteza Mardani, Thorsten Kurth, David Hall, Zongyi Li, Kamyar Azizzadenesheli, Pedram Hassanzadeh, Karthik Kashinath, Animashree Anandkumar, 2022: FourCastNet: A Global Data-driven High-resolution Weather Model using Adaptive Fourier Neural Operators, arXiv, https://doi.org/10.48550/arXiv.2202.11214

Zhong Xiaohui, Lei Chen, Jun Liu, Chensen Lin, Yuan Qi, and Hao Li, 2023: FuXi-Extreme: Improving extreme rainfall and wind forecasts with diffusion model, arXiv, https://doi.org/10.48550/arXiv.2310.19822